\documentstyle[twoside,fleqn,espcrc2]{article}

\newcommand{\be}{\begin{equation}}
\newcommand{\ee}{\end{equation}}
\newcommand{\ba}{\begin{eqnarray}}
\newcommand{\ea}{\end{eqnarray}}
\newcommand{\Tr}{{\rm Tr}}

\newcommand{\AmS}{{\protect\the\textfont2
  A\kern-.1667em\lower.5ex\hbox{M}\kern-.125emS}}

\title{High-energy quark-quark scattering and the eikonal approximation}

\author{E. Meggiolaro 
        \address{Dipartimento di Fisica
        dell'Universit\`a and I.N.F.N., Sezione di Pisa, \\ 
        I--56100 Pisa, Italy.}
        }

\begin{document}

\begin{abstract}
The high--energy quark--quark scattering amplitude is calculated first in 
the case of scalar QCD, using Fradkin's approach to derive the scalar quark 
propagator in an external gluon field and computing it in the eikonal 
approximation. The results are then extended to the case of ``real'' (i.e., 
fermion) QCD. The high--energy quark--quark scattering amplitude turns out to 
be described by the expectation value of two lightlike Wilson lines, running 
along the classical trajectories of the two colliding particles. 
Interesting analytic properties of the high--energy quark--quark scattering 
amplitude can be derived, going from Minkowskian to Euclidean theory: they 
could open the possibility of evaluating the high--energy scattering amplitude 
directly on the lattice.
\end{abstract}

\maketitle

It is known that for a class of {\it soft} high--energy scattering processes,
i.e., elastic scattering processes at high squared energies $s$ in 
the center of mass and small squared transferred momentum $t$ (that is
$s \to \infty$ and $|t| \ll s$, let us say $|t| \le 1~{\rm GeV}^2$),
QCD perturbation theory cannot be safely applied, since $t$ is too small.
Elaborate procedures for summing perturbative contributions have been 
developed \cite{Cheng-Wu-book} \cite{Lipatov}, even if the results are not 
able to explain the most relevant phenomena.

A non--perturbative analysis, based on QCD, of these high--energy 
scattering processes was performed by Nachtmann in \cite{Nachtmann91}.
He studied the $s$ dependence of the quark--quark (and quark--antiquark) 
scattering amplitude by analytical means, using a functional integral 
approach and an eikonal approximation to the solution of the Dirac equation 
in the presence of an external non--Abelian gauge field.

In a previous paper \cite{Meggiolaro96} we proposed an approach to 
high--energy quark--quark (and quark--antiquark) scattering, 
based on a first--quantized path--integral 
description of quantum field theory developed by Fradkin in the early 1960's
\cite{Fradkin}. In this approach one obtains convenient expressions for the 
full and truncated--connected scalar propagators in an external 
(gravitational, electromagnetic, etc.) field and the eikonal approximation 
can be easily recovered in the relevant limit. Knowing the 
truncated--connected propagators, one can then extract, in the manner of 
Lehmann, Symanzik, and Zimmermann (LSZ), the scattering matrix elements 
in the framework of a functional integral approach. 
This method was originally adopted
in \cite{Veneziano} in order to study the four--dimensional Planckian--energy 
gravitational scattering. 

One starts, for simplicity, with the case of {\it scalar} QCD, i.e., 
the case of a spin--0 quark (described by the scalar field $\phi$) coupled 
to a non--Abelian gauge field $A^\mu \equiv A^\mu_a T_a$, $T_a$ 
($a=1, \ldots ,N_c^2 - 1$) being the generators of the Lie algebra of the 
colour group $SU(N_c)$.

In the Fradkin's approach \cite{Fradkin} one writes the 
propagator for the scalar field $\phi$ as a functional integral of the 
first quantized theory \cite{Feynman}.

The Green function (Feynman propagator) for the scalar field $\phi$ in the 
external metric $g_{\mu\nu}$ and in the external non--Abelian gauge field 
$A^\mu$ admits a representation in terms of a functional integral over 
trajectories \cite{Fradkin} \cite{Veneziano}:
\ba
\lefteqn{
G(y,x|g,A) = \displaystyle\int_0^\infty d\nu \int [\sqrt{-g(X)} dX^\mu] }
\nonumber \\
& & \times \delta (x-X(0)) \delta (y-X(\nu)) \times
\nonumber \\
& & {P} \exp \left[ i \displaystyle\int_0^\nu d\tau \, L(X,\dot{X}) \right] ~,
\ea
where
\ba
\lefteqn{
L(X,\dot{X}) = } \nonumber \\
& & {1 \over 2} (g_{\mu\nu} \dot{X}^\mu \dot{X}^\nu \cdot {\bf 1}
+ 2g A_\mu \dot{X}^\mu -m^2 \cdot {\bf 1}) ~.
\ea
The matrix ${\bf 1}$ is the $N_c \times N_c$ unity matrix in the colour space
and ``$P$'' means ``{\it path--ordered}'' along the ``histories''
$X^\mu (\tau)$. Moreover $g(X) \equiv 
\det (g_{\mu\nu} (X))$, so that $g(X) = -1$ in a flat space--time.
The propagator (1) is already connected, i.e., vacuum diagrams have 
been already divided out. Yet, in order to derive the scattering matrix 
elements following the LSZ approach, we need to know the on--shell
{\it truncated--connected} Green functions, which are obtained from the 
connected Green functions by removing the external legs calculated 
on--shell.
In Ref. \cite{Veneziano} it is shown how to compute the truncated--connected 
propagator $\tilde{G}_t (p,p'|A)$, starting from Eq. (1), in the so--called
{\it eikonal} approximation, which is valid in the case of scattering 
particles with very high energy ($E \equiv p^0 \simeq |{\bf p}| \gg m$) and 
small transferred momentum $q \equiv p' - p$ (i.e., $|t| \ll E$, where
$t = q^2$). In this limit the functional integral can be evaluated by means 
of a saddle--point approximation in which the classical trajectory is 
computed to the lowest non--trivial order, that is a straight line (in the 
Minkowski space--time) described by a classical particle (of mass $m$) 
moving with four--momentum $p^\mu$:
\be
X^\mu_0 (\nu) = b^\mu + {p^\mu \over m} \nu ~.
\ee
One thus finds, after properly generalizing the results of \cite{Veneziano} 
to the case of an external non--Abelian gauge field, the following 
expression for the truncated--connected propagator in the eikonal 
approximation:
\ba
\lefteqn{
\tilde{G}_t (p,p'|A) \simeq 2E \displaystyle\int d^3 b \, e^{iqb} \times }
\nonumber \\
& & \left( {P} \exp \left[ -ig \displaystyle\int_{-\infty}^{+\infty}
A_\mu (b + p\tau) p^\mu d\tau \right] - {\bf 1} \right)
\ea
As said before, $q = p' - p$ is the transferred momentum. In ``$d^3 b$'' 
one must not include the component of $b^\mu$ which is parallel to $p^\mu$.
For example, if $p^\mu \simeq p'^\mu \simeq (E,E,0,0)$, one has that
$p_+ \simeq p'_+ \simeq 2E$ and $p_- \simeq p'_- \simeq 0$, where the 
general notation $A_+ \equiv A^0 + A^1$, $ A_- \equiv A^0 - A^1$
has been used for a given four--vector $A^\mu$.
In this case one must take: $d^3 b = d^2 {\bf b}_t db_-$, where
${\bf b}_t = (b_y, b_z)$ is the component of $b^\mu$ in the {\it 
transverse} plane $(y,z)$ (while the two light--cone coordinates,
$x_+ = t + x$ and $x_- = t - x$, are sometimes 
called {\it longitudinal} coordinates).

Knowing the truncated--connected propagators, one can then extract, using the 
reduction equations of LSZ in the framework of a functional integral approach, 
the scattering amplitude of two high--energy spin--0 quarks, in the limit 
$s \to \infty$ and $t \ll s$.
The results can then be easily extended to the case of physical interest,
i.e., the case of ``real'' (fermion) QCD.
 
The quark--quark scattering amplitude, at high squared 
energies $s$ in the center of mass and small squared transferred momentum $t$ 
(that is $s \to \infty$ and $|t| \ll s$, let us say $|t| \le 1~{\rm GeV}^2$), 
turns out to be described by the expectation value of two lightlike Wilson 
lines, running along the classical trajectories of the two colliding particles
\cite{Nachtmann91} \cite{Meggiolaro96}.

In the center--of--mass reference system (c.m.s.), taking for example the 
initial trajectories of the two quarks along the $x^1$--axis, 
the scattering amplitude has the following form 
[explicitly indicating the color indices ($i,j, \ldots$)
and the spin indices ($\alpha, \beta, \ldots$) of the quarks]
\ba
\lefteqn{
M_{fi} = \langle \psi_{i\alpha}(p'_1) \psi_{k\gamma}(p'_2) | M | 
\psi_{j\beta}(p_1) \psi_{l\delta}(p_2) \rangle } \nonumber \\
& & \mathop{\sim}_{s \to \infty}
-{i \over Z_\psi^2} \cdot \delta_{\alpha\beta} \delta_{\gamma\delta}
\cdot 2s 
\displaystyle\int d^2 {\bf z}_t e^{i {\bf q} \cdot {\bf z}_t}
\times \nonumber \\
& & \times \langle [ W_1 (z_t) - {\bf 1} ]_{ij} [ W_2 (0) - {\bf 1} ]_{kl} 
\rangle ~,
\ea
where $q = (0,0,{\bf q})$, with $t = q^2 = -{\bf q}^2$, is the tranferred 
four--momentum and $z_t = (0,0,{\bf z}_t)$, with ${\bf z}_t = (z^2,z^3)$, 
is the distance between the two trajectories in the {\it transverse} plane.
The expectation 
value $\langle f(A) \rangle$ is the average of $f(A)$ in the sense of the 
functional integration over the gluon field $A^\mu$ (including also the 
determinant of the fermion matrix, i.e., $\det[i\gamma^\mu D_\mu - m]$,
where $D^\mu = \partial^\mu + ig A^\mu$ is the covariant derivative)
\cite{Nachtmann91} \cite{Meggiolaro96}.
The two lightlike Wilson lines $W_1 (z_t)$ and $W_2 (0)$ in Eq. (5) are 
defined as
\ba
\lefteqn{W_1 (z_t) =
{P} \exp \left[ -ig \displaystyle\int_{-\infty}^{+\infty}
A_\mu (z_t + p_1 \tau) p_1^\mu d\tau \right] ~, }
\nonumber \\
\lefteqn{W_2 (0) =
{P} \exp \left[ -ig \displaystyle\int_{-\infty}^{+\infty}
A_\mu (p_2 \tau) p_2^\mu d\tau \right] ~, }
\ea
where $P$ stands for ``{\it path ordering}'' and $A_\mu = A_\mu^a T^a$;
$p_1 \simeq (E,E,{\bf 0}_t)$ and $p_2 \simeq (E,-E,{\bf 0}_t)$ are the 
initial four--momenta of the two quarks.
The space--time configuration of these two Wilson lines is shown in Fig. 1.

\newcommand{\wilson}[1]{
\begin{figure}
\begin{center}
\setlength{\unitlength}{1.00mm}
\raisebox{-50\unitlength}
{\mbox{\begin{picture}(80,45)(-35,-30)
\thicklines
\put(-22,22){\line(1,-1){41}}
\put(-15,15){\vector(-1,1){1}}
\put(16,23){\line(-1,-1){42}}
\put(9,16){\vector(1,1){1}}
\put(0,0){\vector(-2,-1){13}}
\thinlines
\put(-8,0){\line(1,0){35}}
\put(8,4){\line(-2,-1){35}}
\put(0,-8){\line(0,1){35}}
\put(27,0){\vector(1,0){1}}
\put(0,27){\vector(0,1){1}}
\put(-13,20){\makebox(0,0){$W_2$}}
\put(18,20){\makebox(0,0){$W_1$}}
\put(-6,-7){\makebox(0,0){$z_t$}}
\put(25,-2){\makebox(0,0){$x$}}
\put(-2,25){\makebox(0,0){$t$}}
\end{picture}}}
\parbox{7.5cm}{\small #1}
\end{center}
\end{figure}}

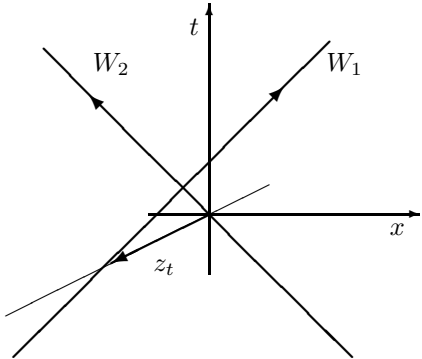
\wilson{{\bf Fig.~1.} The space--time configuration of the two \\
lightlike Wilson lines $W_1$ and $W_2$ entering in the \\
expression (5) for the high--energy quark--quark \\
elastic scattering amplitude.}

Finally, $Z_\psi$ in Eq. (5) is the fermion--field renormalization constant, 
which can be written in the eikonal approximation as \cite{Nachtmann91} 
\be
Z_\psi \simeq 
{1 \over N_c} \langle \Tr [ W_1 (0) ] \rangle
= {1 \over N_c} \langle \Tr [ W_2 (0) ] \rangle ~.
\ee
In a perfectly analogous way one can also derive the high--energy 
scattering amplitude in the case of the Abelian group $U(1)$ (QED).
The resulting amplitude is equal to Eq. (5), with the only obvious 
difference being that now the Wilson lines $W_1$ and $W_2$ are 
functionals of the Abelian field $A^\mu$ (so they are not matrices).
Thanks to the simple form of the Abelian theory (in particular to the 
absence of self--interactions among the vector fields), it turns out that 
it is possible to explicitly evaluate (at least in the {\it quenched} 
approximation) the expectation value of the two Wilson lines 
\cite{Meggiolaro96}: one finally recovers the well--known result for the 
eikonal amplitude of the high--energy scattering in QED \cite{Cheng-Wu} 
\cite{Abarbanel-Itzykson} \cite{Jackiw}.

In what follows, we shall deal with the quantity
\ba
\lefteqn{
g_{M (ij,kl)} (s; ~t) \equiv {1 \over Z_\psi^2}
\displaystyle\int d^2 {\bf z}_t e^{i {\bf q} \cdot {\bf z}_t} \times }
\nonumber \\
& & \times \langle [ W_1 (z_t) - {\bf 1} ]_{ij} [ W_2 (0) - {\bf 1} ]_{kl} 
\rangle ~,
\ea
in terms of which the scattering amplitude can be written as
\ba
\lefteqn{
M_{fi} = \langle \psi_{i\alpha}(p'_1) \psi_{k\gamma}(p'_2) | M | 
\psi_{j\beta}(p_1) \psi_{l\delta}(p_2) \rangle }
\nonumber \\
& & \mathop{\sim}_{s \to \infty}
-i \cdot 2s \cdot \delta_{\alpha\beta} \delta_{\gamma\delta}
\cdot g_{M (ij,kl)} (s; ~t) ~.
\ea
The quantity $g_{M (ij,kl)} (s; ~t)$ depends not only on $t = -{\bf q}^2$,
but also on $s$. In fact, as was pointed out by Verlinde and Verlinde in 
\cite{Verlinde}, it is a singular limit to take the Wilson lines in
(8) exactly lightlike. A way to regularize this sort of ``infrared'' 
divergence (so called because it essentially comes from the limit $m \to 0$,
where $m$ is the quark mass) consists in letting each line
have a small timelike component, so that they coincide with the classical 
trajectories for quarks with a finite mass $m$ (see also Ref. \cite{zfp97} 
for a discussion about this point). In other words, one first evaluates
the quantity $g_{M (ij,kl)} (\beta; ~t)$ for two Wilson lines along the 
trajectories of two quarks (with mass $m$) moving with velocity $\beta$ and 
$-\beta$ ($0 <  \beta <  1$) in the $x^1$--direction.
This is equivalent to consider two infinite Wilson lines forming a 
certain (finite) hyperbolic angle $\chi$ in Minkowski space--time.
Then, to obtain the correct high--energy scattering amplitude, one has to 
perform the limit $\beta \to 1$, that is $\chi \to \infty$, in the 
expression for $g_{M (ij,kl)} (\beta; ~t)$:
\ba
\lefteqn{
M_{fi} = \langle \psi_{i\alpha}(p'_1) \psi_{k\gamma}(p'_2) | M | 
\psi_{j\beta}(p_1) \psi_{l\delta}(p_2) \rangle }
\nonumber \\
& & \mathop{\sim}_{s \to \infty}
-i \cdot 2s \cdot \delta_{\alpha\beta} \delta_{\gamma\delta}
\cdot g_{M (ij,kl)} (\beta \to 1; ~t)
\ea
Proceeding in this way one obtains 
a $\ln s$ dependence of the amplitude, as expected from ordinary 
perturbation theory and as confirmed by the experiments on hadron--hadron 
scattering processes \cite{Cheng-Wu-book} \cite{Lipatov}.

For deriving the dependence on $s$ one exploits the fact that both $\beta$ 
and $\chi$ are dependent on $s$. In fact, from $E = m/\sqrt{1 - \beta^2}$ and 
from $s = 4E^2$, one immediately finds that
\be
\beta = \sqrt{ 1 - {4 m^2 \over s} } ~.
\ee
By inverting this equation and using the relation $\beta = \tanh \psi$, 
where $\psi$ is the hyperbolic angle [in the plane $(x^0,x^1)$] of the 
trajectory of $W_1$, we derive that
\be
s = 4 m^2 \cosh^2 \psi = 2 m^2 ( \cosh \chi + 1 ) ~.
\ee
Therefore, in the high--energy limit $s \to 
\infty$ (or $\beta \to 1$), the hyperbolic angle $\chi = 2\psi$ is 
essentially equal to the logarithm of $s$ (for a finite non--zero quark 
mass $m$):
\be
\chi = 2\psi \mathop{\sim}_{s \to \infty} \ln s ~.
\ee
In Sect. 3 of Ref. \cite{zfp97} we have followed this procedure to explicitly 
evaluate the second member of (10) up to the fourth order in the expansion 
in the renormalized coupling constant $g_R$: the results so derived are
in agreement with those obtained by applying 
ordinary perturbation theory to evaluate the scattering amplitude up to the 
order $O(g_R^4)$ \cite{Cheng-Wu-book} \cite{Lipatov}.

The direct evaluation of the expectation value (8) is a 
highly non--trivial matter and it is strictly connected with the
ultraviolet properties of Wilson--line operators \cite{Arefeva80}.
Some non--perturbative approaches for the calculation of (8) have been 
proposed in Refs. \cite{Arefeva94} and \cite{Dosch}.
At the moment, the only non--perturbative numerical estimate of (8), 
which can be found in the literature, is that of Ref. \cite{Dosch} (where
it has been generalized to the case of hadron--hadron scattering): it has 
been obtained in the framework of the stochastic vacuum model (SVM).

In some recent papers \cite{zfp97} \cite{hep-th/9702186} we have proposed a 
new approach, which consists in adapting the scattering amplitude to the 
Euclidean world: this approach could open the way for the direct evaluation 
of the scattering amplitude on the lattice. More explicitly, we have shown 
that the expectation value of two infinite Wilson lines,
forming a certain hyperbolic angle in Minkowski space--time, and the 
expectation value of two infinite Euclidean Wilson lines, forming a certain 
angle in Euclidean four--space, are connected by an analytic continuation in 
the angular variables. 

Let us consider the following quantity, defined in
Minkowski space--time:
\ba
\lefteqn{
g_M (p_1, p_2; ~t) = {1 \over Z_W^2}
\displaystyle\int d^2 {\bf z}_t e^{i {\bf q} \cdot {\bf z}_t} \times }
\nonumber \\
& & \times \langle [ W_1 (z_t) - {\bf 1} ]_{ij} [ W_2 (0) - {\bf 1} ]_{kl} 
\rangle ~,
\ea
where $p_1$ and $p_2$ are the four--momenta [lying (for example) in the plane 
$(x^0,x^1)$], which define the trajectories of the two Wilson lines $W_1$ and 
$W_2$ ($A_\mu = A_\mu^a T^a$ and $m$ is the fermion mass):
\ba
\lefteqn{W_1 (z_t) \equiv
{P} \exp \left[ -ig \displaystyle\int_{-\infty}^{+\infty}
A_\mu (z_t + {p_1 \over m} \tau) {p_1^\mu \over m} d\tau \right] ~,}
\nonumber \\
\lefteqn{W_2 (0) \equiv
{P} \exp \left[ -ig \displaystyle\int_{-\infty}^{+\infty}
A_\mu ({p_2 \over m} \tau) {p_2^\mu \over m} d\tau \right] ~.}
\ea
$Z_W$ in Eq. (14) is defined as
\be
Z_W \equiv 
{1 \over N_c} \langle \Tr [ W_1 (0) ] \rangle = {1 \over N_c}
\langle \Tr [ W_2 (0) ] \rangle ~.
\ee
This is a sort of Wilson--line's renormalization constant:
as shown in Ref. \cite{Nachtmann91}, $Z_W$ coincides with the fermion 
renormalization constant $Z_\psi$ in the eikonal approximation.

By virtue of the Lorentz symmetry, we can define $p_1$ and $p_2$ in the 
c.m.s. of the two particles, moving with speed $\beta$ and $-\beta$ along 
the $x^1$--direction: the hyperbolic angle $\chi$ [in the plane $(x^0,x^1)$] 
between the two trajectories of $W_1$ and $W_2$ is thus given by
$\chi = 2~ {\rm arctanh} \beta$.

In an analogous way, we can consider the following quantity, defined
in Euclidean space--time:
\ba
\lefteqn{
g_E (p_{1E}, p_{2E}; ~t) = {1 \over Z_{W E}^2}
\displaystyle\int d^2 {\bf z}_t e^{i {\bf q} \cdot {\bf z}_t} \times }
\nonumber \\
& & \times \langle [ W_{1 E} (z_{t E}) - {\bf 1} ]_{ij} [ W_{2 E} (0) - 
{\bf 1} ]_{kl} \rangle_E ~,
\ea
where $z_{t E} = (z_1, z_2, z_3, z_4) = (0, {\bf z}_t, 0)$ and
$q_E = (0, {\bf q}, 0)$ (so that: $q_E^2 = {\bf q}^2 = -t$).
The expectation value $\langle \ldots \rangle_E$ must be intended now as a 
functional integration with respect to the gauge variable $A^{(E)}_\mu = 
A^{(E)a}_\mu T^a$ in the Euclidean theory.
The Euclidean four--vectors $p_{1E}$ and $p_{2E}$ [lying (for example) in the 
plane $(x_1,x_4)$] define the trajectories of the two Euclidean Wilson lines 
$W_{1 E}$ and $W_{2 E}$.
$Z_{W E}$ in Eq. (17) is defined analogously to $Z_W$ in Eq. (16):
\be
Z_{W E} \equiv 
{1 \over N_c} \langle \Tr [ W_{1 E} (0) ] \rangle = {1 \over N_c}
\langle \Tr [ W_{2 E} (0) ] \rangle ~.
\ee
Denoting by $\theta$ the Euclidean angle in the plane $(x_{1},x_{4})$ between 
the trajectories of the two Euclidean Wilson lines $W_{1E}$ and $W_{2E}$, 
one finally finds the following relation between the amplitudes 
$g_M (\chi; ~t)$ and $g_E (\theta; ~t)$ \cite{hep-th/9702186}:
\ba
\lefteqn{g_M (\chi; ~t)
\mathop{\longrightarrow}_{\chi \to i\theta}
g_M (i\theta; ~t) = g_E (\theta; ~t) ~;}
\nonumber \\
\lefteqn{g_E (\theta; ~t)
\mathop{\longrightarrow}_{\theta \to -i\chi}
g_E (-i\chi; ~t) = g_M (\chi; ~t) ~.}
\ea
This relation of analytic continuation has been proven in Ref. \cite{zfp97} 
for an Abelian gauge theory (QED) in the so--called {\it quenched} 
approximation and for a non--Abelian gauge theory (QCD) up to the fourth 
order in the renormalized coupling constant in perturbation theory.
In Ref. \cite{hep-th/9702186}, we have generalized the results of Ref. 
\cite{zfp97} giving the rigorous proof of the above--mentioned relation of 
analytic continuation for a non--Abelian gauge theory with gauge group 
$SU(N_c)$ [as well as for an Abelian gauge theory (QED)]. 
The approach adopted in Ref. \cite{zfp97} consisted in explicitly evaluating 
the amplitudes $g_M$ and $g_E$ in some given approximation (such as the
{\it quenched} approximation) or up to some order in perturbation theory and 
in finally comparing the two expressions so obtained. Instead, in 
Ref. \cite{hep-th/9702186} we have given a general proof, which essentially 
exploits the relation between the gluonic Green functions in the two theories.

Therefore, it is possible to reconstruct the high--energy scattering 
amplitude by evaluating a correlation of two infinite Wilson lines forming a 
certain angle $\theta$ in Euclidean four--space, then by continuing this 
quantity in the angular variable, $\theta \to -i \chi$, where $\chi$ is 
the hyperbolic angle between the two Wilson lines in Minkowski 
space--time, and finally by performing the limit $\chi \to \infty$ (i.e.,
$\beta \to 1$). In fact, the high--energy scattering amplitude is given by
\ba
\lefteqn{
M_{fi} = \langle \psi_{i\alpha}(p'_1) \psi_{k\gamma}(p'_2) | M | 
\psi_{j\beta}(p_1) \psi_{l\delta}(p_2) \rangle } \nonumber \\
& & \mathop{\sim}_{s \to \infty}
-i \cdot 2s \cdot \delta_{\alpha\beta} \delta_{\gamma\delta}
\cdot g_M (\chi \to \infty; ~t) ~.
\ea
The quantity $g_M (\chi; ~t)$, defined by Eq. (14) in the Minkowski world, is 
linked to the corresponding quantity $g_E (\theta; ~t)$, defined by Eq. (17) 
in the Euclidean world, by the analytic continuation (19) in the 
angular variables.
The important thing to note here is that the quantity $g_E (\theta; ~t)$, 
defined in the Euclidean world, may be computed non perturbatively 
by well--known and well--established techniques, for example
by means of the formulation of the theory on the lattice or using the 
SVM \cite{Dosch}. 
In all cases, once one has obtained the quantity $g_E (\theta; ~t)$, one 
still has to perform an analytic continuation in the angular variable
$\theta \to -i \chi$, and finally one has to extrapolate to the limit 
$\chi \to \infty$ (i.e., $\beta \to 1$). For deriving the dependence on $s$
one exploits the fact that both $\beta$ and $\chi$ 
are dependent on $s$, according to Eqs. (11), (12) and (13).

Of course, the most interesting results are 
expected from an {\it exact} non perturbative approach, for example by 
directly computing $g_E (\theta; ~t)$ on the lattice: a considerable 
progress could be achieved along this direction in the near future.

\newpage
\noindent
{\bf DISCUSSIONS}

\vspace{0.5cm}
\noindent
{\bf L.N. Lipatov}, INP, St. Petersburg (Russia)

\noindent
{\it What is the relation of your work with the BFKL results (LLA for 
high--energy scattering)? In 1986 the quark--quark scattering amplitudes 
with the elastic and quasi--elastic unitarity were constructed. Are they 
related with your eikonal amplitude?}

\vspace{0.3cm}
\noindent
{\bf E. Meggiolaro}

\noindent
{\it It is known that ``soft'' hadronic collisions at high energy can be 
described by the Pomeron--exchange model. However, a more basic 
understanding of Pomeron exchange in terms of the underlying quark--gluon 
dynamics (QCD) remains an outstanding open problem. Most theorists now 
agree that the properties of the Pomeron can be somehow derived in QCD from 
multi--gluon exchange and the physical picture commonly accepted is the one 
where the Pomeron couples to single partons: yet, it is still unclear how 
to develop a reliable calculational framework based on this idea. The first 
calculations of the near--forward parton--parton scattering amplitudes were 
done to the lowest orders of perturbation theory (PT) in the leading 
logarithmic approximation (LLA), or using the elastic and quasi--elastic 
unitarity. In particular, the high--energy asymptotics of the scattering 
amplitude was found in the LLA, using the results of one--loop calculations 
in order to resum the leading logarithmic corrections to all oders of PT.
However, in spite of the considerable progress so achieved, there was no 
regular way to resum and control non--leading logarithmic corrections to 
the scattering amplitude. 

\noindent
The approach described in my work enables us to 
take into account systematically these non--leading corrections. It starts 
from the first principles of QCD and derives the high--energy quark--quark 
scattering amplitude in a fully non--perturbative way, using a functional 
integral approach and the so--called eikonal approximation to write the 
truncated--connected quark propagator in an external gluon field. A 
comparison of this approach to the standard PT results has been performed 
up to the one--loop level (fourth order in the renormalized coupling 
constant): the two approaches give the same results.
An explicit comparison with the BFKL Pomeron model is not yet available. It 
would be an extremely important result to derive the features of the 
Pomeron model starting from my non--perturbative approach, based on first 
principles of QCD: this is matter for future work.}

\end{document}